\documentclass[sigconf]{acmart}

\AtBeginDocument{%
  \providecommand\BibTeX{{%
    \normalfont B\kern-0.5em{\scshape i\kern-0.25em b}\kern-0.8em\TeX}}}

\acmConference[UCC ’23]{UCC ’23: IEEE/ACM 16th International Conference on Utility and Cloud Computing
  }{December 04--07, 2023}{Taormina (Messina), Italy}
\begin{document}

%%
%% The "title" command has an optional parameter,
%% allowing the author to define a "short title" to be used in page headers.LLD:
\title{ A Last-Level Defense for Application
Integrity and Confidentiality}

\author{Gabriel P. Fernandez}
\orcid{1234-5678-9012}
\affiliation{%
  \institution{Technische Universit\"at Dresden}
  \streetaddress{Fakultat Informatik}
  \city{Dresden}
  \country{Germany}
%  \postcode{01062}
}
\email{gabriel.pereira_fernandez@tu-dresden.de}

\author{Andrey Brito}
\affiliation{%
  \institution{Federal University of Campina Grande}
  \streetaddress{Fakultat Informatik}
  \city{Campina Grande}
  \country{Brazil}
%  \postcode{}
}
\email{andrey@computacao.ufcg.edu.br}

\author{Ardhi Putra Pratama Hartono}
\affiliation{%
  \institution{Technische Universit\"at Dresden}
  \streetaddress{Fakultat Informatik}
  \city{Dresden}
  \country{Germany}
%  \postcode{}
}
\email{ardhi_pp.hartono@tu-dresden.de}

\author{Muhammad Usama Sardar}
\affiliation{%
  \institution{Technische Universit\"at Dresden}
  \streetaddress{Fakultat Informatik}
  \city{Dresden}
  \country{Germany}
%  \postcode{}
}
\email{muhammad_usama.sardar@tu-dresden.de}

\author{Christof Fetzer}
\affiliation{%
  \institution{Technische Universit\"at Dresden}
  \streetaddress{Fakultat Informatik}
  \city{Dresden}
  \country{Germany}
%  \postcode{}
}
\email{christof.fetzer@tu-dresden.de}

\begin{abstract}
Our objective is to protect the integrity and confidentiality of applications operating in untrusted environments. Trusted Execution Environments (TEEs) are not a panacea. Hardware TEEs fail to protect applications against Sybil, Fork and Rollback Attacks and, consequently, fail to preserve the consistency and integrity of applications. We introduce a novel system, LLD, that enforces the integrity and consistency of applications in a transparent and scalable fashion. Our solution augments TEEs with instantiation control and rollback protection.
Instantiation control, enforced with TEE-supported leases, mitigates Sybil/Fork Attacks without incurring the high costs of solving crypto-puzzles. Our rollback detection mechanism does not need excessive replication, nor does it sacrifice durability. We show that implementing these functionalities in the LLD runtime automatically protects applications and services such as a popular DBMS.
\end{abstract}

%%
%% The code below is generated by the tool at http://dl.acm.org/ccs.cfm.
%% Please copy and paste the code instead of the example below.
%%


%%
%% Keywords. The author(s) should pick words that accurately describe
%% the work being presented. Separate the keywords with commas.

%% A "teaser" image appears between the author and affiliation
%% information and the body of the document, and typically spans the
%% page.

%%
%% This command processes the author and affiliation and title
%% information and builds the first part of the formatted document.
\maketitle

\section{Introduction}

In this work, we introduce \textit{Last-Level Defense (LLD)} -- a TEE-based multi-level architecture to execute trusted applications on enclaves that resist Rollback, Fork, and Sybil Attacks in untrusted environments as basic TEE protection does not cover these classes of attack. In this work, we strive to protect applications with minimal assumptions over their operations and have complete transparency for clients and the application binaries themselves.

In compromised environments, attackers can force illegal executions, changing code in memory or disk (e.g., usurpation attacks). Attackers can also start additional instances to interfere with replication and agreement protocols, leading to inconsistencies of replicated data or routing clients maliciously to spurious service instances to disrupt data consistency (e.g., Sybil Attacks \cite{douceur2002sybil}). They can modify time perception (e.g., to use old Time-based One Time Password -- TOTP \cite{m2011totp} -- authentication codes or expired certificates). Moreover, they can roll back storage to circumvent access revocations and undo undesired actions (e.g., Repudiation Attacks \cite{feng2011enhancing}).
 
Our objective is to protect the integrity and confidentiality of application code and data. Hardening the complete software stack is infeasible as it might be controlled by an untrusted entity like a cloud or service provider using closed-source components. This work builds upon existing cloud orchestration \cite{tosatto2015container, weerasiri2017taxonomy, bousselmi2014cloud} and confidential computing \cite{russinovich2021toward, rashid2020rise,  sardar2021confidential,  sardar2022understanding} work to propose a \textit{Last-Level Defense} for application services. We focus on ensuring the integrity of services in a holistic way, emphasizing the Sybil/Fork Attack vectors (which we collectively refer to as \textit{Instantiation Attacks}), not covered by current TEEs and for which existing solutions \cite{Hartono2021brofy,meena2013survey} have serious limitations. \textit{LLD} supports basic confidentiality provided by \textit{Trusted Execution Environments} (TEE).
Ensuring more advanced confidentiality requires a side-channel protection mechanism -- an orthogonal issue outside this paper’s scope. 

While TEEs such as Intel SGX and AMD SEV provide basic integrity protection, they do not protect against Instantiation Attacks, i.e., the malicious starting of multiple service instances to obtain control over a replica group protocol or manipulate volatile state by controlling traffic to/from instantiated replicas.
Applications that require mutual trust among nodes or a consistent soft state among replicas could be susceptible to these Instantiation Attacks. Clients may rely on trusted configuration, authentication, and attestation mechanisms to identify legitimate server enclaves, but instances of the same enclave will withstand these precautions.% Likewise, a naive solution based on configuration to limit the number of instances of an enclave cannot overcome the Fail-stop failure model -- it cannot know if the running enclave instance has crashed in order to allow the instantiation of a new one.

Permissionless blockchains like Bitcoin use crypto-puzzles to mitigate Sybil Attacks~\cite{hellani2018blockchain}. Solving these puzzles introduces high overhead, high energy consumption, and a substantial monetary cost of running blockchain infrastructures. While blockchain is a revolutionary solution for distributed trusted applications of large dimensions, its threat model assumes the authenticity of most participants. Distributed applications that require trust will not always have the scale to discourage a \textit{51\% Attack}. Even newer blockchain models, such as \textit{Proof-of-Stake}, which do away with the need for computationally intensive Proof-of-Work puzzles, are susceptible to a majority takeover.

We contribute with the three components of \textit{LLD}. The \textit{LLD Runtime} ensures that enclave applications cannot alter their states without authorization. The \textit{LLD Lease Service} ensures enclave applications running with the \textit{LLD Runtime} work as \textit{single-instance} services, i.e., at any time, at most one instance runs, by evaluating their \textit{execution leases}. Our novel defense mechanism’s core is an integrity- and rollback-protected, replicated state database (\textit{LLD Store}). The \textit{LLD Store} keeps leases and guarantees their integrity and durability. Another of our contributions is the novel leader election, a protocol to extend Instantiation-attack protection to \textit{LLD Store} nodes against an unbounded number of spurious instances. 
With \textit{LLD}, off-the-shelf applications a high level of integrity and confidentiality, even in the presence of an adversary that controls the infrastructure, without compromising existing cloud orchestration tools ability to maintain availability.

We organize the paper as follows. Section~\ref{sec:background} details the problem and introduces the technologies use and a motivating application.
Section~\ref{sec:threatmodel} details the expected capabilities and limitations of the adversary.
Section~\ref{sec:approach} summarizes the LLD approach. Section~\ref{sec:monotonic} describes how monotonic counters enable Instantiation Attack tolerance and rollback protection into core services. Then, Section~\ref{sec:time} describes a trusted time model that supports our trusted lease system and how it protects running applications from Instantiation Attacks. We depict our formal evaluation of the local leader protocol that protects our \textit{LLD Store} instances from Sybil Attacks in Section~\ref{sec:formal}. Finally, we evaluate the performance of the proposed approach in protecting an unmodified database server in Section~\ref{sec:eval} and compare our contributions with related work in Section~\ref{sec:rwork}.

\section{Background}
\label{sec:background}

This section reviews the attacks we focus on and then discusses the core technologies that enable our approach.

\subsection{Rollbacks, Sybil/Fork Attacks and their impact on applications}
\label{sec:attacks}

Rollback Attacks~\cite{meena2013survey} create a record of an application’s previous state and use it to overwrite the current state of that application. Although less powerful than an Usurpation Attack, i.e. utilizing high privilege to tamper with application data, a Rollback Attack may undo a legitimate change in the application state to achieve the attacker’s goal. Many applications’ most critical integrity assumption is their ability to keep their most recent state unadulterated. Rollback attacks are effective at disrupting that property.

The Fork Attack (also known as Split View Attack) consists of creating replicas of an application and manipulating the traffic towards the replica group. After manipulating the result, the attacker can terminate the additional replicas. Then, the attacker exploits control over routing to have different values on different replica groups.

For example, an application could be cloned and a targeted transaction could be sent to the clone. After the clone’s state is updated and the operation is committed and acknowledged, the adversary terminates the clone. In this context, a Fork Attack equates to a Rollback Attack. 

This attack works because, in principle, faithful replicas succeed in any authentication mechanisms that the storage servers may have since they are identical copies of the initial replica group. Similarly, TEE’s enclave identification mechanisms will fail to prevent this attack.

In our interpretation of enclave-data-targeting Fork Attacks, a global view of the enclave data is not necessarily available. Solutions proposed in other works like SUNDR and Depot \cite{mahajan2011depot, li2004secure} successfully create protocols that track the internal enclave state. However, they presume the application state is always summarizable by, for example, a hash function. While the case study of file systems fits those solutions’ requirements, it is impractical for applications in general to be able to represent their complete internal state. Furthermore, our transparency requirement prevents solutions from modifying the application. In those solutions the applications actively generates hashes after every meaningful state change. 

We define Sybil Attacks in the context of enclave-based distributed protocols. The adversary can spawn an indefinite number of participants in a peer-to-peer protocol to provoke desired (by the adversary) collective states. In the context of TEE’s, enclave-defined traits, like the MRENCLAVE in the case of Intel SGX, are the equivalent to \textit{identity} in systems where enclaves are attested as a form of authentication. In protocols with an unbounded number of participants (such as blockchains and many conventional consensus protocols), the ability to instantiate enclaves from a certain image indefinitely grants the instantiator the power to subvert collective decisions, even if they cannot control instances’ internal states.
Arbitrary cloning could enable the attacker to multiply a minority state, that is more aligned with his goal, while isolating other replicas when necessary.
Limiting the adversary’s ability to instantiate a certain enclave is an effective way to prevent this attack. 

One Configuration-based straw-man solution to this sort of Sybil Attack would be to leverage the TEE feature of attesting physical nodes and only allow certain nodes to participate. That solution would pin instantiated \textit{client} replicas to specific physical nodes, which is significantly detrimental to scalability, elasticity, migration, and reliability -- the ability of a participant to influence the protocol would be tied to the availability of a physical node.

In our Threat Model, these attacks are equivalent. Restraining the adversary from freely instantiating a certain enclave will avoid both attacks. Therefore, we will collectively refer to them as \textit{Instantiation Attacks}.

\subsection{Dqlite}
\label{subsec:dqlite}

\textit{Dqlite} is a distributed Database Management System (DBMS) based on the broadly used \textit{SQLite} embedded DBMS. It enables running replicas that share the database state to improve availability. Dqlite runs an instance of the \textit{Raft}~\cite{ongaro2014search} algorithm to tackle the problem of the Fail-Stop failure model. In \textit{Dqlite}, the state of the database plays the role of the \textit{Raft} log, thus inheriting its robust persistence properties.

\subsection{TEE Stack}

Our Trusted Execution Environment platform of choice is Intel’s \textit{Software Guard Extensions} (SGX)\cite{costan2016intel}. SGX is a hardware extension that allows applications to ensure confidentiality and integrity despite compromises in higher privilege levels.

SGX trusted execution environment abstractions are called \textit{Enclaves}. Enclaves are hardware-protected memory regions that keep users’ code and data. A hardware-assisted memory encryption engine decrypts that memory region when its pages are moved to cache.

We use \textit{SCONE} as our SGX framework. It optimizes SGX’s thread model, decreasing the costs of transitions between the enclave and the native. It also dramatically simplifies the development model. SCONE adds runtime essentials in the enclave that act as some kernel functions. Its \textit{Configuration and Attestation Service (CAS)} identifies, authenticates, and configures authorized enclaves in cloud environments.

\section{Threat Model}
\label{sec:threatmodel}

We assume a powerful adversary with complete control over the entire software stack in a cloud-like environment. This adversary controls the cloud management system and can launch and delay the instantiation or termination of virtual machines and containers, including those containing enclaves. The same applies to the operating systems instantiated in these resources. They can also terminate containers and virtual machines. 

The adversary also controls the storage, being able to erase and overwrite any content, snapshot any data, safely store it, and replace freshly written data. The adversary can also delay read and write operations to non-volatile memory.

The adversary can intercept any packets sent through the cloud's internal network. The adversary can isolate components or drop, change, delay, or replay packets. Nevertheless, the adversary cannot decrypt messages to which they do not possess the keys.

We assume the adversary cannot compromise the TEE. Our threat model excludes side-channel attacks on enclave memory. Hence, SGX and its dependencies should work correctly. Other works, such as BROFY~\cite{Hartono2021brofy}, approach the correctness of the TEE, which is an orthogonal issue to this work. Monotonic Counters are authenticated and cannot be virtualized. They cannot be overflown and its monotonic invariant is respected, i.e., they cannot be decremented. We consider the adversary to be unwilling to provoke prolonged unavailability, as this can be externally detected. Thus, even though we expect them to delay progress in specific moments, they will eventually allow the whole system to progress, as Service Level Agreements must be respected. Therefore, any performance degradation and forced unavailability attacks (such as \textit{Denial of Service}) will eventually be followed by periods in which the system can progress.

\section{Approach}
\label{sec:approach}

The \textit{LLD} architecture focuses on Instantiation Attack resistance, which we enforce even if an adversary has root access to the hosts, VMs, containers, and infrastructure orchestration. Our approach is to execute applications automatically inside individual TEEs and provide these with the \textit{LLD Runtime}, which enforces instantiation control. In this section, we layout the problems with a naive configuration-based solution and go through \textit{LLD}'s components and their roles in the system.

\subsection{Strawman Solution}

With a simple wrapper around the application code that takes control after initialization, an enclave could contact a configuration server that verifies whether that specific application already has another running enclave. If so, it would order the new instance to shut itself down. This would prevent multiple enclave instances of a certain application from running simultaneously, hindering Instantiation Attacks. As long as that server had some way to communicate with the running instance, it could check for its liveness whenever a new enclave got initialized.

If an instance is unable to communicate with that configuration server, however, the server would have to either allow the new instance to run or to wait for the old instance to respond. In the case that it allowed the new instance to run, there is a chance that the old instance was still alive, it had only gotten slow or deliberately paused -- an instance of the Fail-stop failure model. In that case the single instance invariant would be broken. In our threat model the adversary controls the network and may delay messages to force the configuration server to allow a new instantiation. 
If, however, it always waits for the running instance to respond and the application actually crashes, the configuration will never allow a new instance to run, completely stopping the application from recovering. We conclude that there must be a time-based mechanism that allows applications to keep running. Like in \cite{trach2020t}, we call that abstraction a \textit{lease} -- a period during which an enclave is allowed to execute. 

Furthermore, there must be a solution in-place that prevents the configuration server itself from suffering an Instantiation Attack. Consequently, if there is a single instance of that configuration server, there must be ways to prevent it from being unavailable, since throughout its down time no new applications maybe instantiated or may recover from crashes. 

\subsection{Design And Components}
\label{sec:design}

We overcome the challenges faced by the strawman solution by controlling the application runtime and forcing it to regularly renew its execution lease, providing a fast database-cache frontend and keeping a durable and highly available database backend for those \textit{leases}. In this section we introduce each of these components and the execution flow of the system.

{\bf LLD Runtime:}
Instantiation Attacks depend on the ability to create multiple instances of an enclave. To prevent this, on the application enclave side, we use the \textit{LLD Runtime}, an extension to SCONE's runtime, to verify the validity of its execution \textit{lease}. The \textit{LLD Runtime} runs within the same enclave as the target application and periodically takes over control of execution. When executing, the \textit{LLD Runtime} verifies whether it is still allowed to run by checking the \textit{LLD Lease Service} via network. If not, the \textit{LLD Runtime} terminates the enclave.

The \textit{LLD Runtime} executes every 1 second or whenever a system call that may externalize the internal enclave state may happen. The \textit{LLD Runtime} intercepts system call requests and executes its code instead, a functionality already available in SCONE. We perform this interception because the only way for an enclave internal state to reach a client is via a system call. Thus, if multiple instances of an enclave exist simultaneously, they cannot externalize their state without verifying whether they can continue running by the \textit{LLD Lease Service}. In other words, if a running instance that lost its lease tries to reply to a request or change the persistent application state, we force it to check its lease beforehand, which causes it to terminate itself without succeeding.

{\bf LLD Lease Service:}
To keep track of all running instances for the different applications, we introduce the \textit{LLD Lease Service}, a network-accessible, trusted component comprised of a cached database that authorizes or disavows \textit{LLD Runtime} requests to start or continue running. Its records, which describe our \textit{leases}, are application- and instance-specific time periods allotted to enclave instances. A particular instance can only renew them until its \textit{lease execution period} expires. Once it is, another instance may obtain it. If an enclave instance A fails the deadline for its \textit{lease} and the \textit{LLD Lease Service} has already received a lease request by a new instance B, then A will terminate, and B will keep on executing. Similarly, if a new instance B is instantiated and tries to claim the \textit{lease} owned by A before it expires, B will terminate, and A will continue executing.

A lease request is comprised of an instance ID and a timestamp that the \textit{LLD Runtime} fetches from a trusted time source, representing the instant that instance makes the request. When the \textit{LLD Lease Service} receives it, it compares the timestamp to its database entry for that lease. The entry contains the expiration timestamp for the lease and its owner's instance ID. If the lease request is valid, i.e., the request timestamp is earlier than the entry's expiry timestamp or the instance ID's match, the \textit{LLD Lease Service} renews the lease by updating the expiry timestamp. A typical lease length would be in the order of seconds.

Note that the \textit{LLD Lease Service} in-memory database itself may be the target of Instantiation Attacks. On read-only operations, it replies to clients directly, without accessing the persistent database, the \textit{LLD Store}. On write operations, it verifies the integrity between its state and that of the persistent database and prevents inconsistencies, as further explained in Section~\ref{sec:caching}.

{\bf LLD Store:} 
Lease entries are loaded to the \textit{LLD Lease Service} from the \textit{LLD Store}, a distributed, integrity- and rollback-protected database. It uses a cache image of the loaded state to do read operations over the lease data. When receiving lease updates, which require writes, the \textit{LLD Lease Service} updates the \textit{LLD Store} and reloads its cache. We use a monotonic-counter-based protocol to ensure the \textit{LLD Store} is also resistant to Instantiation Attacks, as we detail in Section \ref{sec:monotonic}.

\begin{figure}[htb]
  \centering
  \includegraphics[width=7cm]{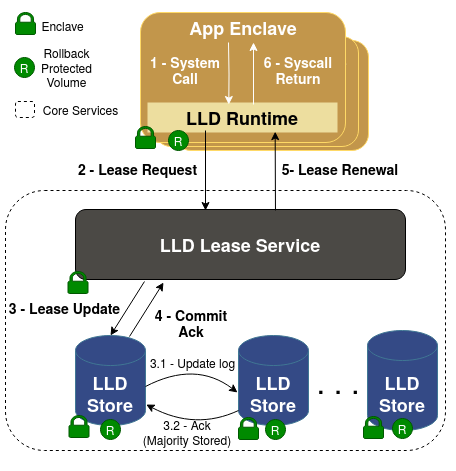}

  \caption{Architecture: Components support LLD's lease update critical path: a state altering system call may trigger a lease renew and be intercepted by the \textit{LLD Runtime}. If the lease is renewed, the \textit{LLD Runtime} returns control to the application.}
  
  \label{fig:tl_architecture}
\end{figure}

Unlike enclaves running with the \textit{LLD Runtime}, \textit{LLD Store} cluster instances do not need to be able to scale or be migrated. Its consensus protocol requires only that over half of the replicas are available. Because of these liberties, they can be pinned down to specific nodes through configuration and work as a source of anti-instantiation attack trust. However, client applications will not have these restrictions, they can be migrated or reinstantiated freely, provided they can still communicate withe the \textit{LLD Lease Service}. Through the \textit{LLD Lease Service}, we extrapolate \textit{LLD Store}'s trust to the \textit{LLD Runtime} and consequently to enclave applications. A minimal cluster of \textit{LLD} core services, as illustrated in \ref{fig:tl_architecture}, may handle thousands of application enclaves with \textit{LLD Runtimes}.

Enclave configuration ties the \textit{LLD Lease Service} to a specific cluster instance of the \textit{LLD Store}. Enclave instances running with the \textit{LLD Runtime} will only authenticate \textit{LLD Lease Services} tied to that cluster instance. This way, the \textit{LLD Store} works as a trust anchor, which prevents the whole architecture from suffering an Instantiation Attack.

\section{Rollback and Instantiation Attack Tolerance for Core Services}
\label{sec:monotonic}

We employ the modules and procedures described further in the section to secure the \textit{LLD Runtime} and the \textit{Lease Server}.

\subsection{Trusted Store}
\textit{LLD} utilizes a \textit{replicated store system}, in the form of a certain number of replicas participating in a cluster, as seen in Figure~\ref{fig:lld-store-implementation}. The cluster runs a consensus algorithm on top of a database management system. 

\textit{LLD}'s lease provisioning service keeps the leases' state and verifies their validity by comparing the deadline of that application's entry to the timestamp in the requesting application. The correctness of that operation ensures the service's ability to prevent unwanted instantiation of an application and, thus, Instantiation Attacks. For that, it requires the Store to have the following properties:

\begin{itemize}
  \item \textbf{Integrity:} If an adversary is capable of changing the contents of the storage, lease information could be altered in such a way as to allow multiple instances of the same application, which permits Instantiation Attacks.
  
  \item \textbf{Consistency:} In our threat model, instances can be crashed or made inaccessible at any moment. Lease data retrieved from all replicas must be consistent with one another, or an adversary may target specific instances to force the \textit{Lease Server} to return some stale lease status.
  
  \item \textbf{Rollback Resistance:} Provided we can prevent integrity breaches, the possibility remains that the adversary restores a previously valid state. 
  Exploiting this, an adversary could send stale lease data to the Lease Server.
\end{itemize}

\subsubsection{Dqlite}

In the context of the \textit{LLD Store},
\textit{Dqlite} provides hard Consistency guarantees. Given a \textit{Dqlite} cluster with $2\cdot N$ participants, the \textit{LLD Lease Server} only reads majority-stored, i.e., committed entries to at least \(N + 1\) replicas (seen on the bottom of Figure \ref{fig:lld-store-implementation}). If availability falls below the \(N + 1\) simple majority threshold, no leader will be elected, and no entries will be majority stored. Therefore, the nodes will not answer any new requests.

\subsubsection{File System Protection}

\textit{SCONE} 's \textit{FSPF} (\textit{File System Protection File}) is a mechanism we leverage to assure file integrity. It consists of an encrypted file that only a certain \textit{SCONE} enclave can access. Data written to this file system branch is encrypted, and the aforementioned \textit{CAS} distributes its keys.

We utilize this \textit{SCONE} feature on the directory containing the \textit{SQlite} file utilized by \textit{Dqlite} on each \textit{Dqlite} replica (represented by the $R$ in Figure \ref{fig:lld-store-implementation}). With a transparently encrypted state, we enforce integrity and confidentiality without making any further changes to the \textit{Dqlite} application.

\subsubsection{Monotonic Counter Supported Updates}

\begin{figure}[htb]
  \centering
  \includegraphics[width=7cm]{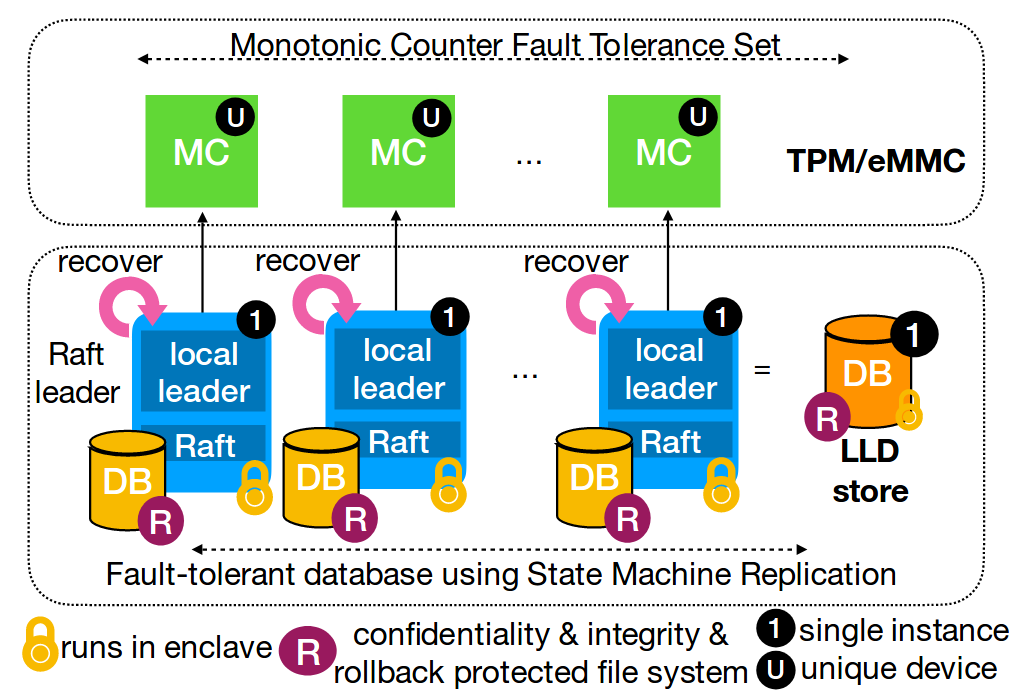}
  \caption{Implementation of \textit{LLD Store}: the multiple Dqlite replicas synchronized via Raft comprise the \textit{LLD Store}}.
  \label{fig:lld-store-implementation}
\end{figure}

On top of the \textit{FSPF} provided encryption of the \textit{Dqlite} files and its timestamp checks, we add monotonic counter updates to the writes (seen on the top of Figure~\ref{fig:lld-store-implementation}).

An adversary could still roll back the file state to a previously valid one. Nevertheless, adding a monotonic counter incremented with each write would invalidate the check of the counter. The counter to which we compare the one in the file gets stored on a monotonic counter device, available in each \textit{Dqlite} node.

\subsubsection{Local Leader Election}

One way to disrupt \textit{Dqlite} provided Consistency is to attack its \textit{Raft} algorithm by performing Instantiation Attacks on the replica nodes themselves, for example, by replicating instances that reorder operations according to one's goals (as discussed in Section \ref{sec:attacks}). To prevent that, we utilize a mechanism that leverages the serializable nature of monotonic counter updates. The protocol, described in Section \ref{sec:local_election}, does not allow two instances of the \textit{Dqlite} enclave to run on the node served by a monotonic counter device, effectively preventing Instantiation Attacks on the local level.
With that solution in place, instantiating other replicas with the same attested \textit{Dqlite} binary would not assist the adversary in the attack, as they would terminate themselves or the running instance. Using a modified binary instead would fail the attestation performed with \textit{CAS} when starting.

This solution protects the \textit{Lease Server} store, extrapolating its safe storage to the client applications. Note that an Instantiation attack on the \textit{Local Leader Election} can only be made via rollback of the written state of the election, hence the employment of trusted monotonic counters.

\section{Trusted Time and Trusted Leases}
\label{sec:time}

As the name suggests, our Trusted Leases are bound to expire at a particular time. From that comes the need for clock readings that can be trusted and consistent, within a certain margin, across the different users of the \textit{Lease Server}. To keep the clocks within said margins, we require them to have bounded \textit{deviation}, i.e., a difference between its value and some collectively agreed upon real-time. To achieve that, we need to establish a maximum clock reading error tolerated for an operation to be considered valid, i.e., the loss in accuracy caused by the time it takes for the clock's reading to complete. The following subsections detail the requirements, model, and resources with which we propose and evaluate \textit{LLD}.

\subsection{Trusted Time Requirements}

When a service running with the \textit{LLD Runtime} reads the time via \texttt{gettimeofday()} and \texttt{clock\_gettime()}, it automatically reads from some trusted \textit{time resource}. To define the properties, let us assume that $t$ denotes
real-time for TAI (Atomic International Time, a.k.a. \texttt{CLOCK\_TAI}). If a thread $T$ calls \texttt{gettimeofday()} at time $t$,  its clock returns the trusted time value $TT_T(t)$. We define the \textit{deviation} of the clock to be $| TT_T(t) - t |$. We require that the maximum clock deviation is $\epsilon$:

\centerline {$\forall t \in Time, \forall T \in Threads:  | TT_T(t) - t | \le \epsilon$ \ \ \ (R1)}

Whereby $\epsilon$ is a configuration parameter that stays constant during the execution of an enclave. This equation implies that the deviation between any two clocks is at most $2\cdot \epsilon$, i.e.:

\centerline {$\forall t \in Time,\forall T, U \in Threads: | TT_T(t) - TT_U(t) | \le 2\cdot\epsilon$}

Some services use time stamps to generate IDs or order events. To support this use case, we require that trusted time is strictly monotonic:

\centerline {$\forall s, t \in Time, \forall T \in Threads: s < t \implies TT_T(s) < TT_T(t)$ \ (R2)}

The clock resolution (a.k.a. clock granularity) is $1\ ns$.

The details for the \emph{Trusted Clock} and its client, implemented in \textit{LLD Runtime}, are outside the scope of this paper, where we focus on defining the Trusted Lease implementation. For now, it suffices to say that, for a certain bound $\epsilon$, R1 is true, and that R2 holds for any two instants $s$ and $t$.

 In the worst case, LLD can query an external trusted time source on every clock read. Nevertheless, much more efficient solutions could be implemented by combining multiple mutually verifying enclave clusters with TSC and a trusted protocol to seed them and eventually re-synchronizing. We place its implementation among our immediate future goals.

\subsection{Trusted Leases Requirements}

We require that a single instance enclave $I$ has an active lease $L$ to continue processing the application program. One expects that, during normal operations, the runtime extends the lease before it expires. However, due to delays or failures, a lease might expire. We automatically check that the lease is still active when the application reads the time: a thread reading the time will block until the lease extension. If \textit{LLD Runtime} cannot extend the lease, e.g., the \textit{LLD Lease Server} has already granted that lease to another instance, $I$ will terminate the enclave. The period in which a instance holds a lease is referred to as its \textit{term}.

Each lease has a unique id. We denote the id of the lease of $T$ as $LID_T(t)$ and the end of the lease time of $T$ at real-time $t$ as $LT_T(t)$. If the lease is expired at $t$, i.e., $LT_T(t) \le TT_T(t)$, we define that $LID_T(t) = \bot$.  

We require that at any point in time, at most one instance has an active lease for a given ID:

\begin{multline*}
    \forall t \in Time, \forall I, J \in Inst: \\  LID_I(t) = LID_J(t) \implies LID_I(t) = \bot \ \ \  (R3)
\end{multline*}

We require that the \textit{LLD Runtime} automatically checks the lease. It can do that at various points in the program. Since we expect leases to be in the seconds range, we determined that we check the lease every time the control returns from an enclave exit. SGX enclaves exit at least every second (as shown in \cite{trach2020t}). Thus, the \textit{LLD Runtime} will read the clock at least once every second when the application is running or when returning from an interruption. Whenever the application reads a clock value $t$, the lease at time $t$ has not yet expired if:

\begin{multline*}
\forall t \in Time, \forall I \in Inst, \forall T \in Threads(I):  \\ LT_I(t) > TT_T(t) \ \ \ \ \ \  (R4)
\end{multline*}

A lease lasts one \textit{Term}, a period during which control is yielded to the application and the \textit{LLD Runtime} does not intercept any system calls.

\subsection{Trusted Lease Implementation}
\label{sec:trusted_lease_impl}

One can specify that, at most, a single instance of a service $S$ can run simultaneously as part of the security policy of $S$. To enforce this single instance requirement of some service $S$, we generate a random lease ID, say, $LID_S$. If a new service instance $I$ of $S$ starts, the \textit{LLD Runtime} gets control first. It generates a random $IID_S$ that uniquely identifies this instance $I$. The \textit{LLD Runtime} will attest $I$ at the \textit{Lease Server}. If no other instance has an active lease for $LID_S$, the \textit{Lease Server} grants the lease $LID_S$ to $I$ until a time $U_S:= LT_I(t)$ and the \textit{CAS} may provision $I$ with secrets. If another instance $I'$ has an active lease, i.e. if $\exists I'ID_{S} | U^{-1}_{S}  > t > U_{S}$, being $U^{-1}_{S}$, the beginning of the \textit{term} $P$, then the \textit{Lease Server} returns an error and the instance $I$ will terminate.

Requirement R4 defines that an instance $I$ can consider a lease $LID_I$ active, i.e., $LID_I(t) \not= \bot$ as long as, for none of its threads, $TT_T(t) \ge LT_I(t)$. When the clock $TT_S$ of the \textit{Lease Server} has reached at least $LT_I(t) + 2\epsilon$, where $LT_I(t)$ is the latest lease granted to $I$, then the \textit{Lease Server} knows that the lease has expired. Nevertheless, it might take another $P$ until the enclave is forced to resynchronize, and all its threads will block until the lease's extension. Hence, the \textit{Lease Server} will not grant a lease to a new instance until $LT_I(t) + 2\epsilon + P$. In this way, we satisfy requirement R3.

\subsubsection{Caching}
\label{sec:caching}

The \textit{Lease Server} has a soft state in the form of the cache copy of the data contained in the persistent storage. This measure does away with the need for querying the database to verify lease states, sensibly improving performance and reducing the blocked time for the client application's \textit{LLD Runtime} thread. 

However, if one instance of the \textit{LLD Lease Server} changed a particular lease id on the database, another \textit{LLD Lease Server} instance's cache would not reflect that change. That difference could cause that other instance to evaluate a lease request for that lease id wrongly.

We do not control instantiation mechanisms in the cloud. Therefore, there is no guarantee that such mechanisms would not create a second \textit{LLD Lease Server}. We preserve the Requirement R3 while maintaining the performance gains achieved with the cache by verifying that transactions done by a certain \textit{LLD Lease Server} instance do not conflict with those of another. We achieve that by utilizing conditional logic queries. We verify that when a lease update gets submitted, the instance id on the \textit{Lease Service} is still the same as in the \textit{LLD Store} . If so, the instance terminates.

If the conditional logic query clause execution fails, indicating that the table contains a different instance id, another \textit{LLD Lease Service} instance must have written updated the \textit{LLD Store}. The \textit{LLD Lease Service} then terminates to avoid further conflict leaving the other instance to continue running.

%%%%%%%%%%%%%%%%%%%%%%%%%%%%%%%%%%%%%
\subsubsection{Local Leader Election}
\label{sec:local_election}

We propose a general-use leader election (\textit{Local Leader Election}) protocol that ensures that there is only one instance of some service at a time on a node. In \textit{LLD}, we employ it to ensure that there is only one \textit{Raft} instance per authorized node. Unlike the leader election in \textit{Raft}, there could be an \textit{unbounded} number of participants competing to become a local leader. Hence, we cannot solve the \textit{Local Leader Election} problem with the help of majority votes like \textit{Raft}. Instead, we use a lease-like mechanism -- without requiring a Lease Server. 

%As opposed to the multi-node \emph{Raft}, the \textit{Local Leader Election} happens in a single \textit{LLD Store} replica, and prevents multiple instances of it to run on the same physical node.

We first introduce the \textit{Local Leader Election} problem more precisely. We distinguish different local elections by assigning each a unique ID $CID$. We select this ID randomly for a single-instance service. 
The \textit{Local Leader Election} ensures that there is at most one local leader at a time for election $CID$. If a local leader fails, we expect an untrusted orchestrator to restart a \textit{candidate}. An orchestrator under the control of an adversary could start multiple concurrent candidates at times when a local leader already exists. 

We specify the \textit{Local Leader Election} in the same way we define trusted leases, i.e., with requirements R3-4. Note, however, that our trusted lease implementation requires the transactional store, which requires \textit{Raft}, which in turn requires that there is at most a single instance per replica running. Therefore, we solve the \textit{Local Leader Election} with the help of monotonic counters. Each monotonic counter has a unique, non-clonable ID. For simplicity, we assume that that is the same ID $CID$ as in the \textit{Raft} election. A monotonic counter, represented by $MC_{CID}$, has the following properties. First, it is monotonic (nondecreasing):

\centerline {$\forall CID \ \forall s, t \in Time:  s < t \implies MC_{CID}(s) \le MC_{CID}(t)$ \ \ (R5)} 
where $MC_{CID}(s)$ and $MC_{CID}(t)$ represent the monotonic counter values at time instances $s$ and $t$, respectively.

Second, reads of the monotonic counters are linearizable, i.e., the value returned by the monotonic counter $MC_{CID}$ was correct at some time between the time $s$ a thread $T$ requests to read $MC_{CID}$, denoted by $readMC_{T,CID}(s)$ and a time $u > s$ a value $V$ was returned $valueMC_{T,s,V}(u)$:

{\center $\forall LID \ \forall s, u \in Time, \forall V: readMC_{T,CID}(s) \wedge valueMC_{T,s,V}(u)$ \newline $\implies \exists t \in [s,u]: MC_{CID}(t)=V $\ (R6)}
% In the implementation, there is only one MC. 

Third, when a write returns that it has successfully incremented the monotonic counter to the desired value, then the monotonic counter has indeed shown this value.

{\center $\forall CID \ \forall s, u \in Time, \forall V: writeMC_{T,CID, V}(s) \wedge valueMC_{T,s,V}(u)$ \newline $\implies \exists t \in [s,u]: MC_{CID}(t)=V $\ (R7)} \\

A service instance $I$ starts as a \textit{candidate} and randomly selects a unique ID, say $ID_I$. We split a monotonic counter value in a pair $(M, ID)$, i.e., an integer counter value and an ID, and define the usual total order on this pair. It reads the monotonic counter value as $V = (C, ID)$. It will then try incrementing the value to $V_N = (C+1, ID_I)$. If the write fails, the candidate's election has failed, and the candidate will terminate. The candidate will eventually become the new local leader if the write succeeds. First, a previous local leader has to be able to detect that a new leader was elected.

\begin{figure}[htb]
  \centering
  \hspace*{-0.1in}
  \includegraphics[width=7cm]{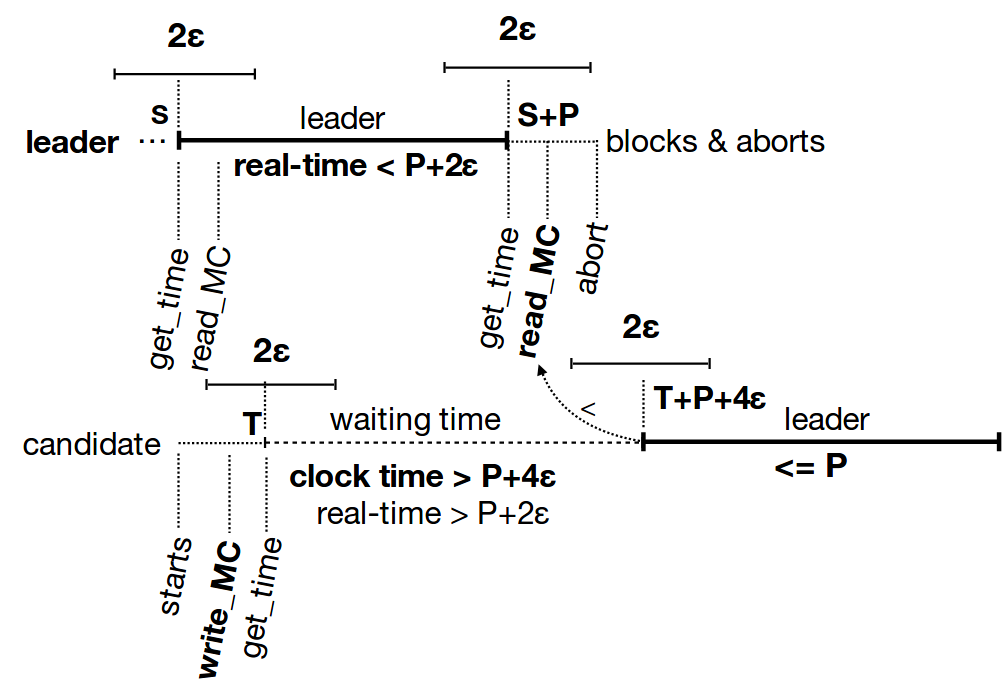}
  \caption{A new local leader needs to wait for at least $P+4\epsilon$ to ensure that a previous leader has learned that it is not the leader anymore .}
\label{fig:local_leader}

\end{figure}

As in the trusted time case, a leader checks every $P$ seconds that it is still the leader. Using trusted time, the measurement error of the length of a period $P$ is in the range of $[-2\epsilon, 2\epsilon]$ since the start and end of an interval can be measured with an error of up to $\pm\epsilon$. Since a new leader and the old leader can error in different directions, the new leader waits for at least $P+4\epsilon$ before it executes its application code. Figure \ref{fig:local_leader} illustrates this.

We expect that the orchestrator starts in almost all cases, a single instance only, i.e., we do not need to optimize for concurrent candidates. We focus on ensuring that there is, at most, one local leader at a time. Moreover, we want to reduce the wear and tear on the monotonic counters by reducing the number of writes to the monotonic counters. Our algorithm writes at most once per candidate. 
We prevent adding new nodes into the cluster by specifying authorized counter devices as part of the enclave configuration. 

\section{Formal Verification of Local Leader Election Protocol}
\label{sec:formal}

We formally verify the local leader protocol to gain confidence in its design, as it is the cornerstone of our architecture. Our general approach is to perform model checking \cite{Clarke2000} of the protocol. The main benefit of model checking is the automatic verification of the properties of interest. If a property does not hold for a given protocol model, it also provides a counter-example for debugging.

Specifically, we use the SPIN model checker \cite{SPIN} mainly because of its performance.

All time-related variables are of type \texttt{unsigned} of parameterized length (represented by \texttt{BITWIDTH}). We also parameterize the term (represented by \texttt{P}). We model the clock reading error as a non-deterministic value selected from the range [$-\epsilon,\epsilon$]. All of these constants \texttt{EPSILON, P, MAXIN} and \texttt{BITWIDTH} are parameterized in the formal model and thus can be provided externally by the user for the desired analysis.  

SPIN uses 2-step semantics, i.e., it evaluates guard and execution in two steps. Therefore, our protocol uses \texttt{atomic} to model the desired behavior. We analyze with the constants $EPSILON=1, MAXIN=3, P=5, BITWIDTH=20$.

In the context of the \textit{Local Leader Election} protocol, the most important property is the safety property, i.e., at any point in time, there must not be two leaders doing the work. We formalize it as the condition that the candidate is a leader and is inside the enclave. Formally, in Linear Temporal Logic (LTL), this can be represented as:
${[] !((leader_x \&\& in_x) \&\& (leader_y \&\& in_y))}$
where $leader_x$ and $leader_y$ represent that the candidates x and y, respectively, are leaders, and $in_x$ and $in_y$ represent that the candidates x and y, respectively, are inside the enclave. ! represents logical negation and $[]$ represents the LTL operator \textit{always}. 

We verify using SPIN version 6.5.1 on Ubuntu 20.04 LTS on an Intel Icelake machine 112-core with a processor base frequency of 3.0 GHz with 512 {GB} RAM. We show the verification statistics in Table \ref{tab:ver-stats}.

\begin{table}[htb]
%\captionsetup[table]{skip=10pt}
\centering

\begin{tabular}{|l|l|}
\hline
Depth             & 1.5x$10^9$ \\ \hline
States            & 4.5x$10^8$ \\ \hline
Transitions       & 1.3x$10^9$ \\ \hline
%Memory used       & 139 GB \\ \hline
Verification time & 53 minutes \\ \hline
 %State-vector size & \mytodo{} \\ \hline

\end{tabular}
\caption{Verification statistics for safety property in the \textit{Local Leader Election} protocol}
\label{tab:ver-stats}

\end{table}
\vspace*{-1cm}
\section{Evaluation}
\label{sec:eval}

We use three Intel NUCs with 4-core Pentium Silver J5005 and SGXv2 processors for the \textit{Dqlite} trusted storage. We use one server with Icelake-SP processors (experimental versions that preceded the released Xeon Scalable 3rd Generation) for the \textit{Lease Server} and clock server, as well as the client application running atop \textit{LLD}. 

There is a variety of option available regarding the choice of monotonic counter device. In our experiments, we utilize a rendition that has a mean verified-write latency of $\sim 40\ ms$.

\subsection{Rollback and Instantiation Attack tolerance on the LLD Store}
\label{sec:dqlite_eval}

Ensuring single instance enclaves requires running the trusted lease services atop rollback- and Sybil-tolerant services. Sybil tolerance for the \textit{LLD Store} requires the \textit{Local Election} mechanism (see Section~\ref{sec:local_election}). This protocol imposes an increase in start-up time and requires that \textit{LLD Store} participants periodically reevaluate their local leadership, interrupting useful workflow. A work thread in the \textit{LLD Store} keeps executing during the leader verification period to minimize this effect. If control is lost, it first checks if the period is over and if it is still the \textit{Local Leader}.

\begin{figure}[htb]
  \centering
  \includegraphics[width=7cm]{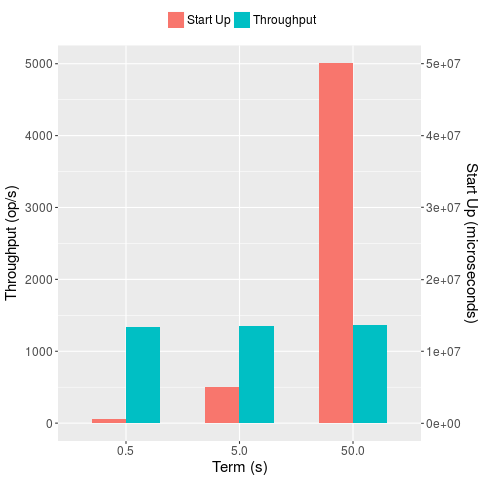}
  \caption{Start-up delay and work lost per period in 3 configurations.}
  \label{fig:lld-store}
  \vspace*{-0.1cm}
\end{figure}

Figure~\ref{fig:lld-store} depicts both the start-up delay and the amount of useful work lost in the periodic leadership checks by the \textit{LLD Store} in a period ($P$) for three different configurations of $P$. As expected, the increase in $P$ and the consequent reduction in the frequency of checks for changes in the leadership had little to no noticeable effect. Notice that the proposed $P$ in \textit{Configuration 1} is lower than the minimum lease period ($1\ s$). With that, we demonstrate that $P = 1\ s$ is an optimal configuration since it comes with virtually no additional cost and provides the best start-up and reduction in Mean-Time-To-Recover in the event of a crash.

\begin{figure}[htb]
  \centering
  \includegraphics[width=6cm]{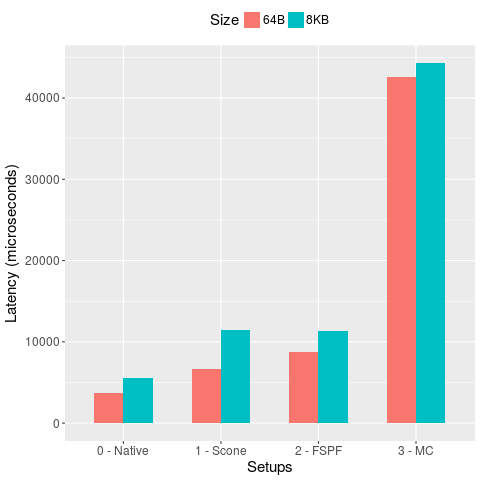}
  \caption{Cost of \textit{LLD} writes with different levels of security.}
  \label{sec:rollback-costs}
  \vspace*{-0.1cm}
\end{figure}

The following experiment evaluates the cost of using an encrypted, integrity-, and rollback-protected storage supported by enclaves and external monotonic counters. For that, we evaluate the performance in operations per second for local writes on two workloads (very short, with 64 bytes, and long, with $8\ KB$) and four scenarios: ($0$) without any protection; ($1$) with only TEE framework-provided integrity protection; ($2$) additionally with previously available storage encryption and integrity protection; and, ($3$) the proposed encryption, integrity, and rollback protection. Figure~\ref{sec:rollback-costs} depicts the results. The cost of the monotonic counter update dominates the latency on setup $3$, as expected. Since our \textit{Dqlite} does them asynchronously, no more than a residual difference between the workload sizes can be noted. Scenario $2$ shows 25\% higher latency than scenario $1$. That difference depicts the cost of encryption relative to the size of the load. However, it completely disappears on the larger buffer size. We attribute that phenomenon to the larger buffer flush policy implemented on the FSPF file system, which is better for larger writes. %The difference compensates for the cost of encryption, with scenario $3$ performing marginally better with that workload.

%%%%%%%%%%%%%%%%%%%%%%%%%%%%%%%%%
\subsection{Extending protection to off-the-shelf services and applications}
\label{sec:mysql_eval}

Next, we evaluate the direct usage of the rollback-protected storage in a \textit{Dqlite} database (used in the implementation of the \textit{LLD Store}) and in a MariaDB server, provisioned by a regular user of the \textit{LLD Runtime}.

\begin{figure}[htb]
  \centering
  \includegraphics[width=7cm]{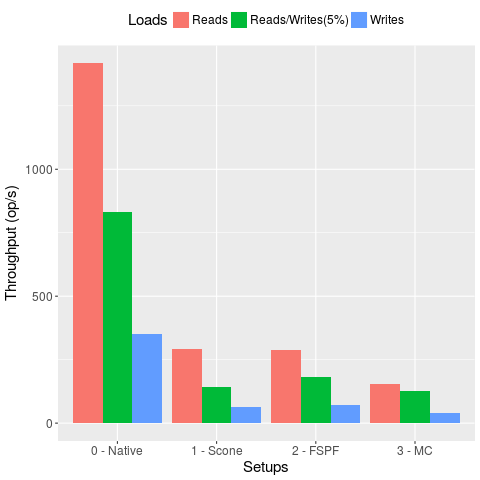}
  \caption{Performance of the \textit{Dqlite} 3-node cluster with increasing layers of protection measures.}
  \label{fig:dqlite}
\end{figure}

Figure~\ref{fig:dqlite} depicts the performance of a \textit{Dqlite} cluster with three load scenarios (read-only, $5\%$ writes, and write-only) and four platform configurations (no protection, regular \textit{SCONE} protection, \textit{SCONE} protection and encryption, and full Instantiation Attack and rollback protection). We set \textit{Dqlite} replication to 3, using the SGXv2 NUCs, and use a custom benchmark, reading and writing 64-byte long records, approximately the size of the structures for lease management.

As we expected, the fact that the majority of the system call occurrences will not trigger the expensive lease update significantly dilutes its impact on performance, as compared with the direct write benchmark previously shown. Scenarios ($1$) and ($2$) have shown similar results. Scenario ($3$) suffers from the accumulated monotonic counter update cost, with a performance loss ranging from 43\% (writes-only) to 49\% (reads-only). 

\begin{figure}[!ht]
  \centering
  \includegraphics[width=7cm, 
 ]{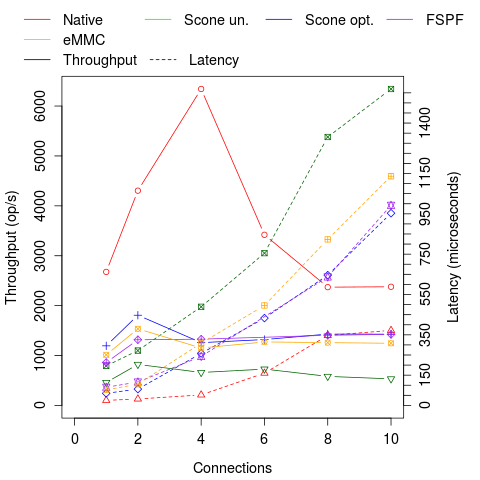}
  \caption{Performance on TPC-C of MariaDB with increasing levels of security.}
  \label{fig:mariadb}
  \vspace*{-0.2cm}
\end{figure}

Figure~\ref{fig:mariadb} depicts the performance of a MariaDB instance under the TPC-C benchmark. We have used five platform configurations (no protection, regular \textit{SCONE} protection, \textit{SCONE} protection with tuned parameters, \textit{SCONE} protection with tuned parameters and \textit{FSPF} encryption, and full Instantiation Attack and rollback protection).

The unoptimized \textit{SCONE} version lacks fine tuning on \textit{SCONE}'s threading parameters. By adjusting sleep times and busy-wait spins for enclave and system call threads, one can better tailor its performance for storage applications, the focus of our evaluation.

Due to the usage of caching, among other more sophisticated database resources present on MariaDB (primarily aimed at reducing I/O), there is little difference between the monotonic-counter-protected setup and the others. We have achieved less than 12\% additional overhead in latency and throughput. As shown in the Figure~\ref{fig:mariadb}, with the more realistic workload the cost of the complete protection package has significantly been amortized, even when evaluated by an I/O intensive benchmark.

%\begin{figure*}[htb]
%  \centering
%  \includegraphics[width=\textwidth]{img/mariaDBTPCHLatency.png}3
%  \caption{Performance on TPC-H of a popular DBMS (MariaDB) with normal execution, basic confidentiality and integrity support, and LLD support.}
%  \label{fig:mariadb-tpch}
%\end{figure*}

\section{Related Work}
\label{sec:rwork}

{\bf Trusted Execution Environments for Distributed Systems}: 
Trusted Execution Environments (TEEs), such as ARM TrustZone~\cite{TrustZone}, IBM SecureBlue++~\cite{SecureBlue1,SecureBlue2}, AMD SEV~\cite{SEV}, and Intel Trusted Domain Extensions (TDX)~\cite{TDX}, enrich the design of complex, yet secure, distributed systems. TEEs offer the ability to ensure the integrity and confidentiality of applications against adversaries with privileged access to components such as the operating system. 

In addition, TEEs support \emph{attestation} mechanisms.
TEEs have been extensively used to improve the performance and resilience of consensus protocols in distributed systems~\cite{ behl2017hybrids, russinovich2019ccf, brandenburger2018blockchain, liu2018scalable}. T-Lease~\cite{trach2020t} strides further to create a low-level trusted lease primitive, although it is inserted in an in-enclave-only threat model.

{\bf Monotonic Counters}: 
Monotonic counters have commonly been used for handling Rollback Attacks in distributed applications. TEEs, e.g., Intel SGX, support a monotonic counter implementation using Intel Management Engine~\cite{ cen2017trusted}. The controllers cannot increment the counters more than $20$ times per second~\cite{palaemon}. To improve the performance, one could keep a version number in an encrypted database and use a hardware-based monotonic counter that keeps track of this version number~\cite{palaemon}. Alternatively, ROTE~\cite{matetic2017rote} uses distributed consensus-based trusted counters to improve the performance of NVMEM-based monotonic counters.

{\bf Fork Attacks}:
In SUNDR \cite{li2004secure}, in the context of trusted storage systems, the authors engage with the problem of Fork Attacks and maintaining Fork consistency. Li et al.'s solution consists of keeping a log of all operations done, with cryptographic signatures provided by the client for each of the operations. After each operation, the system provides clients with a signature of the file system state. That solution resists operation rollbacks and creates inconsistencies between client versions. While SUNDR succeeds in creating disfavourable conditions for Fork attacks, an adversary could still hide operations if communication between clients is infeasible. That solution also requires applications to generate snapshots of internal states, which is natural for file systems but can be awkward and require loss of transparency for soft state applications.

On the same domain, DEPOT~\cite{mahajan2011depot} expands upon it by including a logical clock and update exchanges among the client nodes. While augmenting previous work capabilities, DEPOT suffers from the same limitations. Clients are required to abide by a more rigid protocol that requires state-keeping among each other.

{\bf Sybil Attacks} 
Douceur \cite{douceur2002sybil} demonstrates that only a form of trusted certification could eliminate the possibility of Sybil Attacks but does not provide specific methods. Solutions like the random key predistribution \cite{newsome2004sybil} do not consider a scenario where the key distribution is virtualizable. Similarly, solutions like Cornelli's \cite{cornelli2002implementing} base themselves on verifying network resources, which are virtualizable and thus not trusted in our threat model.

\section{Conclusion}

We propose a platform that implements TEE framework-based functionality and extends it to Sybil/Fork Attacks. That is a difficult task, given that the potential adversary controls all extra-enclave aspects of the operation.
However, we have shown that with monotonic counters, we can create safe protocols for instantiating single-instance services.

We use a distributed database system on these hardened instances to create a lease system capable of extrapolating these properties to all applications hosted in that local area. We argue that this is a  practical, affordable approach to the Sybil/Fork Attack problem.

Finally, we evaluate the cost of employing our solutions and demonstrate that, although the values shown in raw benchmarks of the full protection stack may be high, with a popular benchmark and DBMS, the costs are greatly amortized.

\begin{acks} 
This work was supported by Cluster of Excellence “Centre for Tactile Internet with Human-in-the-Loop” (CeTI) of Technische Universität Dresden with Project ID 390696704, Federal Ministry of Education and Research of Germany in the programme of “Souverän. Digital. Vernetzt.” as project 6G-life with ID: 16KISK001K, and by European Commission through the Horizon Europe Research and Innovation program under Grant Agreement No. 101016577 (AI-SPRINT), 101092644 (NearData), and 101092646 (CloudSkin).
\end{acks}

\bibliographystyle{ACM-Reference-Format}
\bibliography{refs}

%%% -*-BibTeX-*-
%%% Do NOT edit. File created by BibTeX with style
%%% ACM-Reference-Format-Journals [18-Jan-2012].

\begin{thebibliography}{34}

%%% ====================================================================
%%% NOTE TO THE USER: you can override these defaults by providing
%%% customized versions of any of these macros before the \bibliography
%%% command.  Each of them MUST provide its own final punctuation,
%%% except for \shownote{}, \showDOI{}, and \showURL{}.  The latter two
%%% do not use final punctuation, in order to avoid confusing it with
%%% the Web address.
%%%
%%% To suppress output of a particular field, define its macro to expand
%%% to an empty string, or better, \unskip, like this:
%%%
%%% \newcommand{\showDOI}[1]{\unskip}   % LaTeX syntax
%%%
%%% \def \showDOI #1{\unskip}           % plain TeX syntax
%%%
%%% ====================================================================

\ifx \showCODEN    \undefined \def \showCODEN     #1{\unskip}     \fi
\ifx \showDOI      \undefined \def \showDOI       #1{#1}\fi
\ifx \showISBNx    \undefined \def \showISBNx     #1{\unskip}     \fi
\ifx \showISBNxiii \undefined \def \showISBNxiii  #1{\unskip}     \fi
\ifx \showISSN     \undefined \def \showISSN      #1{\unskip}     \fi
\ifx \showLCCN     \undefined \def \showLCCN      #1{\unskip}     \fi
\ifx \shownote     \undefined \def \shownote      #1{#1}          \fi
\ifx \showarticletitle \undefined \def \showarticletitle #1{#1}   \fi
\ifx \showURL      \undefined \def \showURL       {\relax}        \fi
% The following commands are used for tagged output and should be
% invisible to TeX
\providecommand\bibfield[2]{#2}
\providecommand\bibinfo[2]{#2}
\providecommand\natexlab[1]{#1}
\providecommand\showeprint[2][]{arXiv:#2}

\bibitem[Behl et~al\mbox{.}(2017)]%
        {behl2017hybrids}
\bibfield{author}{\bibinfo{person}{Johannes Behl}, \bibinfo{person}{Tobias
  Distler}, {and} \bibinfo{person}{R{\"u}diger Kapitza}.}
  \bibinfo{year}{2017}\natexlab{}.
\newblock \showarticletitle{Hybrids on steroids: SGX-based high performance
  BFT}. In \bibinfo{booktitle}{\emph{Proceedings of the Twelfth European
  Conference on Computer Systems}}.
\newblock


\bibitem[Boivie and Williams(2012)]%
        {SecureBlue2}
\bibfield{author}{\bibinfo{person}{Rick Boivie} {and} \bibinfo{person}{Peter
  Williams}.} \bibinfo{year}{2012}\natexlab{}.
\newblock \showarticletitle{SecureBlue++: CPU support for secure execution}.
\newblock \bibinfo{journal}{\emph{Technical report}} (\bibinfo{year}{2012}).
\newblock


\bibitem[Bousselmi et~al\mbox{.}(2014)]%
        {bousselmi2014cloud}
\bibfield{author}{\bibinfo{person}{Khadija Bousselmi}, \bibinfo{person}{Zaki
  Brahmi}, {and} \bibinfo{person}{Mohamed~Mohsen Gammoudi}.}
  \bibinfo{year}{2014}\natexlab{}.
\newblock \showarticletitle{Cloud services orchestration: A comparative study
  of existing approaches}. In \bibinfo{booktitle}{\emph{2014 28th International
  Conference on Advanced Information Networking and Applications Workshops}}.
  IEEE, \bibinfo{pages}{410--416}.
\newblock


\bibitem[Brandenburger et~al\mbox{.}(2018)]%
        {brandenburger2018blockchain}
\bibfield{author}{\bibinfo{person}{Marcus Brandenburger},
  \bibinfo{person}{Christian Cachin}, \bibinfo{person}{R{\"u}diger Kapitza},
  {and} \bibinfo{person}{Alessandro Sorniotti}.}
  \bibinfo{year}{2018}\natexlab{}.
\newblock \showarticletitle{Blockchain and trusted computing: Problems,
  pitfalls, and a solution for hyperledger fabric}.
\newblock \bibinfo{journal}{\emph{arXiv preprint arXiv:1805.08541}}
  (\bibinfo{year}{2018}).
\newblock


\bibitem[Cen and Zhang(2017)]%
        {cen2017trusted}
\bibfield{author}{\bibinfo{person}{Shanwei Cen} {and} \bibinfo{person}{Bo
  Zhang}.} \bibinfo{year}{2017}\natexlab{}.
\newblock \showarticletitle{Trusted time and monotonic counters with intel
  software guard extensions platform services}.
\newblock \bibinfo{journal}{\emph{Online at: https://software. intel.
  com/sites/default/files/managed/1b/a2/Intel-SGX-Platform-Services. pdf}}
  (\bibinfo{year}{2017}).
\newblock


\bibitem[{Clarke Jr.} et~al\mbox{.}(1999)]%
        {Clarke2000}
\bibfield{author}{\bibinfo{person}{Edmund~M {Clarke Jr.}},
  \bibinfo{person}{Orna Grumberg}, {and} \bibinfo{person}{Doron~A Peled}.}
  \bibinfo{year}{1999}\natexlab{}.
\newblock \bibinfo{booktitle}{\emph{{Model Checking}}}.
\newblock \bibinfo{publisher}{MIT Press}.
\newblock
\showISBNx{0-262-03270-8}


\bibitem[Cornelli et~al\mbox{.}(2002)]%
        {cornelli2002implementing}
\bibfield{author}{\bibinfo{person}{Fabrizio Cornelli}, \bibinfo{person}{Ernesto
  Damiani}, \bibinfo{person}{Sabrina De Capitani~di Vimercati},
  \bibinfo{person}{Stefano Paraboschi}, {and} \bibinfo{person}{Pierangela
  Samarati}.} \bibinfo{year}{2002}\natexlab{}.
\newblock \showarticletitle{Implementing a reputation-aware gnutella servent}.
  In \bibinfo{booktitle}{\emph{International Conference on Research in
  Networking}}. Springer, \bibinfo{pages}{321--334}.
\newblock


\bibitem[Costan and Devadas(2016)]%
        {costan2016intel}
\bibfield{author}{\bibinfo{person}{Victor Costan} {and}
  \bibinfo{person}{Srinivas Devadas}.} \bibinfo{year}{2016}\natexlab{}.
\newblock \showarticletitle{Intel SGX explained}.
\newblock \bibinfo{journal}{\emph{Cryptology ePrint Archive}}
  (\bibinfo{year}{2016}).
\newblock


\bibitem[Douceur(2002)]%
        {douceur2002sybil}
\bibfield{author}{\bibinfo{person}{John~R Douceur}.}
  \bibinfo{year}{2002}\natexlab{}.
\newblock \showarticletitle{The sybil attack}. In
  \bibinfo{booktitle}{\emph{International workshop on peer-to-peer systems}}.
  Springer, \bibinfo{pages}{251--260}.
\newblock


\bibitem[Feng et~al\mbox{.}(2011)]%
        {feng2011enhancing}
\bibfield{author}{\bibinfo{person}{Jun Feng}, \bibinfo{person}{Yu Chen},
  \bibinfo{person}{Douglas Summerville}, \bibinfo{person}{Wei-Shinn Ku}, {and}
  \bibinfo{person}{Zhou Su}.} \bibinfo{year}{2011}\natexlab{}.
\newblock \showarticletitle{Enhancing cloud storage security against roll-back
  attacks with a new fair multi-party non-repudiation protocol}. In
  \bibinfo{booktitle}{\emph{2011 IEEE Consumer Communications and Networking
  Conference (CCNC)}}. IEEE, \bibinfo{pages}{521--522}.
\newblock


\bibitem[Gregor et~al\mbox{.}(2020)]%
        {palaemon}
\bibfield{author}{\bibinfo{person}{Franz Gregor}, \bibinfo{person}{Wojciech
  Ozga}, \bibinfo{person}{S{\'e}bastien Vaucher}, \bibinfo{person}{Rafael
  Pires}, \bibinfo{person}{Sergei Arnautov}, \bibinfo{person}{Andr{\'e}
  Martin}, \bibinfo{person}{Valerio Schiavoni}, \bibinfo{person}{Pascal
  Felber}, \bibinfo{person}{Christof Fetzer}, {et~al\mbox{.}}}
  \bibinfo{year}{2020}\natexlab{}.
\newblock \showarticletitle{Trust management as a service: Enabling trusted
  execution in the face of byzantine stakeholders}. In
  \bibinfo{booktitle}{\emph{2020 50th Annual IEEE/IFIP International Conference
  on Dependable Systems and Networks (DSN)}}.
\newblock


\bibitem[Hartono and Fetzer(2021)]%
        {Hartono2021brofy}
\bibfield{author}{\bibinfo{person}{Ardhi Putra~Pratama Hartono} {and}
  \bibinfo{person}{Christof Fetzer}.} \bibinfo{year}{2021}\natexlab{}.
\newblock \showarticletitle{BROFY: Towards Essential Integrity Protection for
  Microservices}. In \bibinfo{booktitle}{\emph{2021 40th International
  Symposium on Reliable Distributed Systems (SRDS)}}.
  \bibinfo{pages}{154--163}.
\newblock
\urldef\tempurl%
\url{https://doi.org/10.1109/SRDS53918.2021.00024}
\showDOI{\tempurl}


\bibitem[Hellani et~al\mbox{.}(2018)]%
        {hellani2018blockchain}
\bibfield{author}{\bibinfo{person}{Hussein Hellani},
  \bibinfo{person}{Abed~Ellatif Samhat}, \bibinfo{person}{Maroun Chamoun},
  \bibinfo{person}{Hussein El~Ghor}, {and} \bibinfo{person}{Ahmed
  Serhrouchni}.} \bibinfo{year}{2018}\natexlab{}.
\newblock \showarticletitle{On blockchain technology: Overview of bitcoin and
  future insights}. In \bibinfo{booktitle}{\emph{2018 IEEE International
  Multidisciplinary Conference on Engineering Technology (IMCET)}}. IEEE,
  \bibinfo{pages}{1--8}.
\newblock


\bibitem[Holzmann(1997)]%
        {SPIN}
\bibfield{author}{\bibinfo{person}{G.J. Holzmann}.}
  \bibinfo{year}{1997}\natexlab{}.
\newblock \showarticletitle{{The model checker SPIN}}.
\newblock \bibinfo{journal}{\emph{IEEE Transactions on Software Engineering}}
  \bibinfo{volume}{23}, \bibinfo{number}{5} (\bibinfo{date}{may}
  \bibinfo{year}{1997}), \bibinfo{pages}{279--295}.
\newblock
\showISSN{00985589}
\urldef\tempurl%
\url{https://doi.org/10.1109/32.588521}
\showDOI{\tempurl}


\bibitem[{Intel Corporation}(2020)]%
        {TDX}
\bibfield{author}{\bibinfo{person}{{Intel Corporation}}.}
  \bibinfo{year}{2020}\natexlab{}.
\newblock \bibinfo{booktitle}{\emph{An introductory overview of the {Intel}
  {TDX} technology}}.
\newblock \bibinfo{type}{Intel White Paper}. \bibinfo{institution}{{Intel
  Corporation}}.
\newblock


\bibitem[Kaplan et~al\mbox{.}(2016)]%
        {SEV}
\bibfield{author}{\bibinfo{person}{David Kaplan}, \bibinfo{person}{Jeremy
  Powell}, {and} \bibinfo{person}{Tom Woller}.}
  \bibinfo{year}{2016}\natexlab{}.
\newblock \bibinfo{title}{AMD memory encryption}.
\newblock
\newblock
\newblock
\shownote{White paper}.


\bibitem[Li et~al\mbox{.}(2004)]%
        {li2004secure}
\bibfield{author}{\bibinfo{person}{Jinyuan Li}, \bibinfo{person}{Maxwell~N
  Krohn}, \bibinfo{person}{David Mazieres}, {and} \bibinfo{person}{Dennis~E
  Shasha}.} \bibinfo{year}{2004}\natexlab{}.
\newblock \showarticletitle{Secure Untrusted Data Repository (SUNDR).}. In
  \bibinfo{booktitle}{\emph{Osdi}}, Vol.~\bibinfo{volume}{4}.
  \bibinfo{pages}{9--9}.
\newblock


\bibitem[Liu et~al\mbox{.}(2018)]%
        {liu2018scalable}
\bibfield{author}{\bibinfo{person}{Jian Liu}, \bibinfo{person}{Wenting Li},
  \bibinfo{person}{Ghassan~O Karame}, {and} \bibinfo{person}{N Asokan}.}
  \bibinfo{year}{2018}\natexlab{}.
\newblock \showarticletitle{Scalable byzantine consensus via hardware-assisted
  secret sharing}.
\newblock \bibinfo{journal}{\emph{IEEE Trans. Comput.}} (\bibinfo{year}{2018}).
\newblock


\bibitem[Mahajan et~al\mbox{.}(2011)]%
        {mahajan2011depot}
\bibfield{author}{\bibinfo{person}{Prince Mahajan}, \bibinfo{person}{Srinath
  Setty}, \bibinfo{person}{Sangmin Lee}, \bibinfo{person}{Allen Clement},
  \bibinfo{person}{Lorenzo Alvisi}, \bibinfo{person}{Mike Dahlin}, {and}
  \bibinfo{person}{Michael Walfish}.} \bibinfo{year}{2011}\natexlab{}.
\newblock \showarticletitle{Depot: Cloud storage with minimal trust}.
\newblock \bibinfo{journal}{\emph{ACM Transactions on Computer Systems (TOCS)}}
  \bibinfo{volume}{29}, \bibinfo{number}{4} (\bibinfo{year}{2011}),
  \bibinfo{pages}{1--38}.
\newblock


\bibitem[Matetic et~al\mbox{.}(2017)]%
        {matetic2017rote}
\bibfield{author}{\bibinfo{person}{Sinisa Matetic}, \bibinfo{person}{Mansoor
  Ahmed}, \bibinfo{person}{Kari Kostiainen}, \bibinfo{person}{Aritra Dhar},
  \bibinfo{person}{David Sommer}, \bibinfo{person}{Arthur Gervais},
  \bibinfo{person}{Ari Juels}, {and} \bibinfo{person}{Srdjan Capkun}.}
  \bibinfo{year}{2017}\natexlab{}.
\newblock \showarticletitle{$\{$ROTE$\}$: Rollback protection for trusted
  execution}. In \bibinfo{booktitle}{\emph{26th $\{$USENIX$\}$ Security
  Symposium ($\{$USENIX$\}$ Security 17)}}. \bibinfo{pages}{1289--1306}.
\newblock


\bibitem[Meena et~al\mbox{.}(2013)]%
        {meena2013survey}
\bibfield{author}{\bibinfo{person}{Sachin Meena}, \bibinfo{person}{Esther
  Daniel}, {and} \bibinfo{person}{NA Vasanthi}.}
  \bibinfo{year}{2013}\natexlab{}.
\newblock \showarticletitle{Survey on various data integrity attacks in cloud
  environment and the solutions}. In \bibinfo{booktitle}{\emph{2013
  International Conference on Circuits, Power and Computing Technologies
  (ICCPCT)}}. IEEE, \bibinfo{pages}{1076--1081}.
\newblock


\bibitem[M'Raihi et~al\mbox{.}(2011)]%
        {m2011totp}
\bibfield{author}{\bibinfo{person}{David M'Raihi}, \bibinfo{person}{Salah
  Machani}, \bibinfo{person}{Mingliang Pei}, {and} \bibinfo{person}{Johan
  Rydell}.} \bibinfo{year}{2011}\natexlab{}.
\newblock \bibinfo{booktitle}{\emph{Totp: Time-based one-time password
  algorithm}}.
\newblock \bibinfo{type}{{T}echnical {R}eport}.
\newblock


\bibitem[Newsome et~al\mbox{.}(2004)]%
        {newsome2004sybil}
\bibfield{author}{\bibinfo{person}{James Newsome}, \bibinfo{person}{Elaine
  Shi}, \bibinfo{person}{Dawn Song}, {and} \bibinfo{person}{Adrian Perrig}.}
  \bibinfo{year}{2004}\natexlab{}.
\newblock \showarticletitle{The sybil attack in sensor networks: analysis \&
  defenses}. In \bibinfo{booktitle}{\emph{Third international symposium on
  information processing in sensor networks, 2004. IPSN 2004}}. IEEE,
  \bibinfo{pages}{259--268}.
\newblock


\bibitem[Ongaro and Ousterhout(2014)]%
        {ongaro2014search}
\bibfield{author}{\bibinfo{person}{Diego Ongaro} {and} \bibinfo{person}{John
  Ousterhout}.} \bibinfo{year}{2014}\natexlab{}.
\newblock \showarticletitle{In search of an understandable consensus
  algorithm}. In \bibinfo{booktitle}{\emph{2014 USENIX Annual Technical
  Conference (Usenix ATC 14)}}. \bibinfo{pages}{305--319}.
\newblock


\bibitem[Rashid(2020)]%
        {rashid2020rise}
\bibfield{author}{\bibinfo{person}{Fahmida~Y Rashid}.}
  \bibinfo{year}{2020}\natexlab{}.
\newblock \showarticletitle{The rise of confidential computing: Big tech
  companies are adopting a new security model to protect data while it's in
  use-[news]}.
\newblock \bibinfo{journal}{\emph{IEEE Spectrum}} \bibinfo{volume}{57},
  \bibinfo{number}{6} (\bibinfo{year}{2020}), \bibinfo{pages}{8--9}.
\newblock


\bibitem[Russinovich et~al\mbox{.}(2019)]%
        {russinovich2019ccf}
\bibfield{author}{\bibinfo{person}{Mark Russinovich}, \bibinfo{person}{Edward
  Ashton}, \bibinfo{person}{Christine Avanessians}, \bibinfo{person}{Miguel
  Castro}, \bibinfo{person}{Amaury Chamayou}, \bibinfo{person}{Sylvan Clebsch},
  \bibinfo{person}{Manuel Costa}, \bibinfo{person}{C{\'e}dric Fournet},
  \bibinfo{person}{Matthew Kerner}, \bibinfo{person}{Sid Krishna},
  {et~al\mbox{.}}} \bibinfo{year}{2019}\natexlab{}.
\newblock \showarticletitle{CCF: A framework for building confidential
  verifiable replicated services}.
\newblock \bibinfo{journal}{\emph{Technical Report MSR-TR-201916}}
  (\bibinfo{year}{2019}).
\newblock


\bibitem[Russinovich et~al\mbox{.}(2021)]%
        {russinovich2021toward}
\bibfield{author}{\bibinfo{person}{Mark Russinovich}, \bibinfo{person}{Manuel
  Costa}, \bibinfo{person}{C{\'e}dric Fournet}, \bibinfo{person}{David
  Chisnall}, \bibinfo{person}{Antoine Delignat-Lavaud}, \bibinfo{person}{Sylvan
  Clebsch}, \bibinfo{person}{Kapil Vaswani}, {and} \bibinfo{person}{Vikas
  Bhatia}.} \bibinfo{year}{2021}\natexlab{}.
\newblock \showarticletitle{Toward confidential cloud computing}.
\newblock \bibinfo{journal}{\emph{Commun. ACM}} \bibinfo{volume}{64},
  \bibinfo{number}{6} (\bibinfo{year}{2021}), \bibinfo{pages}{54--61}.
\newblock


\bibitem[Sardar(2022)]%
        {sardar2022understanding}
\bibfield{author}{\bibinfo{person}{Muhammad~Usama Sardar}.}
  \bibinfo{year}{2022}\natexlab{}.
\newblock \showarticletitle{Understanding Trust Assumptions for Attestation in
  Confidential Computing}. In \bibinfo{booktitle}{\emph{2022 52nd Annual
  IEEE/IFIP International Conference on Dependable Systems and
  Networks-Supplemental Volume (DSN-S)}}. IEEE, \bibinfo{pages}{49--50}.
\newblock


\bibitem[Sardar and Fetzer(2021)]%
        {sardar2021confidential}
\bibfield{author}{\bibinfo{person}{Muhammad~Usama Sardar} {and}
  \bibinfo{person}{Christof Fetzer}.} \bibinfo{year}{2021}\natexlab{}.
\newblock \showarticletitle{Confidential Computing and Related Technologies: A
  Review}.
\newblock  (\bibinfo{year}{2021}).
\newblock


\bibitem[Tosatto et~al\mbox{.}(2015)]%
        {tosatto2015container}
\bibfield{author}{\bibinfo{person}{Andrea Tosatto}, \bibinfo{person}{Pietro
  Ruiu}, {and} \bibinfo{person}{Antonio Attanasio}.}
  \bibinfo{year}{2015}\natexlab{}.
\newblock \showarticletitle{Container-based orchestration in cloud: state of
  the art and challenges}. In \bibinfo{booktitle}{\emph{2015 Ninth
  international conference on complex, intelligent, and software intensive
  systems}}. IEEE, \bibinfo{pages}{70--75}.
\newblock


\bibitem[Trach et~al\mbox{.}(2020)]%
        {trach2020t}
\bibfield{author}{\bibinfo{person}{Bohdan Trach}, \bibinfo{person}{Rasha
  Faqeh}, \bibinfo{person}{Oleksii Oleksenko}, \bibinfo{person}{Wojciech Ozga},
  \bibinfo{person}{Pramod Bhatotia}, {and} \bibinfo{person}{Christof Fetzer}.}
  \bibinfo{year}{2020}\natexlab{}.
\newblock \showarticletitle{T-Lease: a trusted lease primitive for distributed
  systems}. In \bibinfo{booktitle}{\emph{Proceedings of the 11th ACM Symposium
  on Cloud Computing}}. \bibinfo{pages}{387--400}.
\newblock


\bibitem[Weerasiri et~al\mbox{.}(2017)]%
        {weerasiri2017taxonomy}
\bibfield{author}{\bibinfo{person}{Denis Weerasiri},
  \bibinfo{person}{Moshe~Chai Barukh}, \bibinfo{person}{Boualem Benatallah},
  \bibinfo{person}{Quan~Z Sheng}, {and} \bibinfo{person}{Rajiv Ranjan}.}
  \bibinfo{year}{2017}\natexlab{}.
\newblock \showarticletitle{A taxonomy and survey of cloud resource
  orchestration techniques}.
\newblock \bibinfo{journal}{\emph{ACM Computing Surveys (CSUR)}}
  \bibinfo{volume}{50}, \bibinfo{number}{2} (\bibinfo{year}{2017}),
  \bibinfo{pages}{1--41}.
\newblock


\bibitem[Williams and Boivie(2011)]%
        {SecureBlue1}
\bibfield{author}{\bibinfo{person}{Peter Williams} {and} \bibinfo{person}{Rick
  Boivie}.} \bibinfo{year}{2011}\natexlab{}.
\newblock \showarticletitle{CPU support for secure executables}. In
  \bibinfo{booktitle}{\emph{International Conference on Trust and Trustworthy
  Computing}}.
\newblock


\bibitem[Winter(2008)]%
        {TrustZone}
\bibfield{author}{\bibinfo{person}{Johannes Winter}.}
  \bibinfo{year}{2008}\natexlab{}.
\newblock \showarticletitle{Trusted computing building blocks for embedded
  linux-based ARM trustzone platforms}. In
  \bibinfo{booktitle}{\emph{Proceedings of the 3rd ACM workshop on Scalable
  trusted computing (STC)}}.
\newblock


\end{thebibliography}

%%
%% If your work has an appendix, this is the place to put it.
\appendix

\end{document}